\documentclass[prb,twocolumn,floatfix,showpacs]{revtex4}
\usepackage{psfrag,epsfig} 
\usepackage{graphicx}
\usepackage{amssymb}
\usepackage{amsmath}

\begin{document}
\title{Spin-charge separation in transport through Luttinger liquid rings} 
\author{S.~Friederich} \affiliation{Institut
  f\"ur Theoretische Physik, Universit\"at G\"ottingen, 37077
  G\"ottingen, Germany} 
\author{V.~Meden} \affiliation{Institut f\"ur Theoretische Physik A, RWTH Aachen University, 
52056 Aachen, Germany}

\date{\today}

\begin{abstract}
We investigate how the different velocities characterizing the low-energy spectral 
properties and the low-temperature thermodynamics of one-dimensional correlated 
electron systems (Luttinger liquids) affect the transport properties of  
ring-like conductors. The Luttinger liquid ring is coupled to two noninteracting 
leads and pierced by a magnetic flux. We study the flux dependence of the 
linear conductance. It shows a dip structure which is governed by the
interaction dependent velocities. Our work extends an earlier study which 
was restricted to rather specific choices of the interaction parameters. 
We show that for generic repulsive two-particle interactions the number of 
dips can be estimated from the ratio of the charge current velocity and 
the spin velocity. In addition, we clarify the range of validity 
of the central approximation underlying the earlier study. 
  \end{abstract}
\maketitle     

\section{Introduction}
In the presence of a two-particle interaction the low-temperature thermodynamics 
and the low-energy spectral properties of a wide class of one-dimensional (1d) 
electron systems is described by the Luttinger liquid (LL) 
phenomenology.\cite{Schoenhammer05} One of the characterizing properties of a LL 
is the complete separation of the fundamental collective spin and charge 
excitations. In the low-energy limit and for a fixed number of left and right 
moving electrons the Hamiltonian can be written as the sum 
of two terms describing free bosons for the spin and charge degrees of 
freedom, both having a linear dispersion with velocities $v_s$ and $v_c$. 
Furthermore, in 1d the bosonic degrees of freedom span the entire low-energy 
Hilbert space at fixed particle number. This has to be contrasted to the 
collective spin and 
charge excitations of higher dimensional Fermi liquids which only give a certain 
part of the spectrum. In many theoretical studies the effect of spin-charge 
separation was discussed\cite{Schoenhammer05,Argentinier,VM1,Voit,Uli}   
and there have been several attempts to find experimental indications 
for it.\cite{experiments} 
 
The low-energy physics of a LL is parameterized by four (in general) independent 
and interaction dependent fundamental velocities $v_{N,c}$, $v_{N,s}$, $v_{J,c}$, and 
$v_{J,s}$. The first two are relevant when the total charge or spin are 
changed, while the later two are the velocities characterizing the charge and
spin current. From these, $v_s$ and $v_c$ can be computed as\cite{Schoenhammer05} 
\begin{equation}
\label{defvsc}
v_{c/s}=\sqrt{v_{N,c/s} \;  v_{J,c/s}} \; .
\end{equation}  

Here we are interested in the role of spin-charge separation or more generally 
in the role of the different interaction dependent velocities on transport through 
LLs. There is no specific effect on the linear conductance $G$ of a linear LL wire 
connected to a Fermi liquid source and drain. A different situation occurs if a 
closed LL ring coupled via tunnel barriers to two noninteracting leads is 
considered. This geometry allows for interference and an additional 
parameter can be added to the problem, namely a magnetic flux $\phi$ piercing 
the ring. The linear conductance then becomes a periodic function of $\phi$ 
with a periodicity of one flux quantum $\phi_0$. We here chose units such that 
$\phi_0=2 \pi$.

In Ref.\ \onlinecite{Jagla} it 
was argued that for $v_c/v_s=q/p$, with $q$ and $p$ being ``small'' odd numbers, 
$G(\phi)$ shows characteristic dips whose number for $\phi \in [0, 2 \pi]$ is 
given by $q v_F/v_c=p v_F/v_s$, with 
$v_F$ being the noninteracting Fermi velocity.
The appearance of such dips might thus be interpreted 
as an indication of spin-charge separation.
Restricting the ratio of the charge and spin velocities to such cases, however requires rather
specific interaction parameters which will be hard to realize in experimental systems.
The authors study a continuum model, the Tomonaga-Luttinger model (TLM). 
The central approximation of this work---which can be applied for the 
above given ratio of $v_c$ and 
$v_s$---is related to the idea that an electron can pass the ring only
if the charge and spin can ``recombine'' at the right (and left) 
contact (for details see below). A dip structure of $G(\phi)$  was 
also obtained for a small ring described by a lattice model with 
strong local Coulomb interaction, the 1d $t-J$ model.\cite{Hallberg} 

We here reinvestigate the zero temperature transport through a LL 
ring described by the TLM. We do  
not rely on the above mentioned approximation of Ref.\
\onlinecite{Jagla} and present results for arbitrary repulsive interactions, 
thus being closer to situations which can be realized in experimental setups.  
This extends the earlier study and clarifies the range of applicability 
of the central approximation underlying that work. More precisely, we find 
that the approximation of Ref.\ \onlinecite{Jagla} leads to qualitatively 
correct results only if the smaller of the two numbers, say $p$,\cite{footnote} 
is either $1$, $3$ or $5$, with the refinement that for $p=3$ or $5$
the other number $q$ may not exceed $7$. Furthermore, the important role 
of the charge current velocity $v_{J,c}$ as a factor accompanying the
magnetic flux in all relevant formulas was largely overlooked so far. 
While, according to the approximation made in Ref. \onlinecite{Jagla}, the 
transmission probability exhibits $p \, v_F/v_s=q \, v_F/v_c$ dips over one 
period of the magnetic flux, we find
that the expression $v_{J,c}/v_s$ gives a much better estimate on the
number of dips in almost all cases. It is thus the charge current velocity and not 
the velocity $v_c$ of the bosonic modes which is of relevance for the dip structure
of $G(\phi)$. We also clarify how the two expressions for
the number of dips and their respective partial viabilities are related.

Transport through LL rings weakly coupled to noninteracting leads was also studied 
in Refs.~\onlinecite{Kinaret} and \onlinecite{Mikhail} although with a focus 
different to ours. 

The rest of this paper is organized as follows. To set the stage  
we introduce the TLM and our way to compute the linear
conductance in Sec.\ \ref{model}. The latter involves the one-particle 
Green function of the TLM which we also discuss in this section. 
In Sec.\ \ref{jaglaapprox} 
we describe the approximation of 
Ref.\ \onlinecite{Jagla} and the corresponding results for the flux dependence of 
the linear conductance. Our results for $G(\phi)$ are discussed in 
Sec.\ \ref{results} and compared to the earlier findings of 
Ref.\ \onlinecite{Jagla}. We also comment 
on the relation to the numerical study Ref.\ \onlinecite{Hallberg}. We conclude with 
a summary in Sec.\ \ref{summary}. 
   
\section{Model and methods}
\label{model}

\subsection{The Tomonaga-Luttinger model}

The low-energy physics of 1d correlated metals is governed by 
long-range effective forces. 
Consequently, relevant wave numbers are from relatively narrow momentum intervals
close to the Fermi momenta $\pm k_F$.\cite{Schoenhammer05} 
It is then natural to linearize the free dispersion relation around the 
two Fermi momenta. The electrons are separated into two classes, namely into 
right- and left-moving fermions, respectively. The Hamiltonian then reads 
(in standard second quantized notation)
\begin{align}
H=&H_0+H_{\rm{int}},\\
H_0=&{\sum_k}' \sum_{\alpha,\sigma} \alpha v_F(k-\alpha k_F)c^\dagger_{k,\alpha,\sigma}
c_{k,\alpha,\sigma}, \nonumber \\
H_{\rm{int}}=&\frac{1}{2L}{\sum_{k,k',q}}' \sum_{\sigma,\sigma'}
\sum_{\alpha,\alpha'} \tilde v^{\sigma,\sigma'}_{\alpha,\alpha'}(q)
c^\dagger_{k+q,\alpha,\sigma}c^\dagger_{k'-q,\alpha',\sigma'} \nonumber \\ 
& \times  c_{k',\alpha',\sigma'}
c_{k,\alpha,\sigma} \; , \nonumber
\end{align}
with $\alpha=\pm$ labeling the two branches and $\sigma$ the spin direction.  
The prime at the momentum sums
indicates that the momenta are restricted to a range $2 \Lambda$ 
around the two Fermi points, with the momentum cutoff $\Lambda>0$. 
Since the electrons move in a finite system of length $L$ with periodic 
boundary conditions, the $k$-sum is over discrete values 
$k_n=\frac{2\pi}{L} n$, $n \in {\mathbb Z}$.
The elements of the interaction matrix are conveniently classified according to 
the so-called g-ology convention.\cite{solyom} 
For the TLM the following couplings are relevant:
\begin{eqnarray}
\tilde v^{\sigma,\sigma}_{\alpha,\alpha}(q)=g_{4,\parallel}(q) \; ,
\;\;\;\; \tilde v^{\sigma,-\sigma}_{\alpha,\alpha}(q)=g_{4,\perp}(q) \; ,
\nonumber  \\
\tilde v^{\sigma,\sigma}_{\alpha,-\alpha}(q)=g_{2,\parallel}(q) \; ,
\;\;\;\; \tilde v^{\sigma,-\sigma}_{\alpha,-\alpha}(q)=g_{2,\perp}(q) \; .
\end{eqnarray}
The $g_4$-processes correspond to intra-branch and 
the $g_2$-processes to inter-branch scattering. The
$\parallel$\,-\,index indicates processes involving two electrons 
of parallel spin orientation while $\perp$ stands for scattering of electrons 
with opposite spin.

After normal ordering the operators with respect to the ground state and
sending the momentum cutoff $\Lambda$ to $\infty$ the model can be
solved exactly using bosonization, that is the introduction of 
bosonic creation and annihilation operators.\cite{Schoenhammer05,Herbert} 
The first step towards bosonization 
is to consider the density operators ($q \neq 0$)
\begin{equation}
\rho_{q,\alpha,\sigma}=\sum_kc^\dagger_{k,\alpha,\sigma}c_{k+q,\alpha,\sigma} \; .
\end{equation}
Operators $b^{\dagger}_{q,\sigma}$ and $b_{q,\sigma}$ obeying bosonic commutation 
relations are obtained defining
\begin{equation}
b_{q,\sigma}=\sqrt{\frac{2\pi}{|q|L}}\begin{cases} \rho_{q,+,\sigma} & q>0 
\\ \rho_{q,-,\sigma} & q<0 \end{cases} \; .
\end{equation}
These operators as well as the different couplings $g_2$ and $g_4$ may be mapped 
onto new ones which are no longer associated with the two spin directions 
but with collective charge and spin excitations of the system, respectively ($\nu=2,4$)
\begin{align}
b_{q,c} \equiv \frac{1}{\sqrt{2}} (
b_{q,\uparrow}+b_{q,\downarrow}) \; ,\nonumber\\
b_{q,s} \equiv \frac{1}{\sqrt{2}} (b_{q,\uparrow}-b_{q,\downarrow}) \; ,
\end{align}
\begin{align}
g_{\nu,c}(q) \equiv g_{\nu,\parallel} +g_{\nu,\perp} \; ,\nonumber\\
g_{\nu,s}(q) \equiv g_{\nu,\parallel}-g_{\nu,\perp} \; .
\label{couplingsnew}
\end{align}
We define excess particle number operators ${\mathcal N}_{\alpha,\sigma}$ 
with respect to the ground state (normal ordering). Spin and charge 
operators for a fixed branch index are obtained as the linear combinations 
\begin{equation}
\mathcal N_{\alpha,c/s}=\frac{1}{\sqrt2}\left(\mathcal N_{\alpha,\uparrow} 
\pm\mathcal N_{\alpha,\downarrow}\right) \; , 
\label{ndef}
\end{equation}
charge and spin current operators as ($a=c/s$)
\begin{equation}
\mathcal{J}_a=\mathcal{N}_{+,a}-\mathcal{N}_{-,a}
\label{Jdef}
\end{equation} 
and the total particle number and spin operators as
\begin{equation}
	\mathcal{N}_a=\mathcal{N}_{+,a}+\mathcal{N}_{-,a} \; .
\label{Ndef}
\end{equation}
In terms of these newly defined operators and couplings, the 
Hamiltonian splits up into two commuting parts
\begin{equation}
H^{ \rm TL}=H^{ \rm TL}_c+H^{ \rm TL}_s
\end{equation}
with 
\begin{eqnarray}
H^{\rm TL}_a & = & \sum_{k>0}\left\{ k\Big\lbrack v_F+\frac{g_{4,a}(k)}{2\pi}
\Big\rbrack\left( b^\dagger_{k,a}b_{k,a}+b^\dagger_{-k,a}b_{-k,a} \right)\right.\nonumber 
\\
&& \left. +k\frac{g_{2,a}(k)}{2\pi}\left( b^\dagger_{k,a}b^\dagger_{-k,a}
+b_{-k,a}b_{k,a} \right) \right\} \nonumber \\
&& +\frac{\pi}{2L}\left( v_{J,a}\mathcal{J}_a^2+v_{N,a}\mathcal{N}_a^2 \right) \; .
\label{scs} 
\end{eqnarray} 
The velocities $v_{J,a}$ and $v_{N,a}$ are defined in terms of the 
couplings as
\begin{equation}
v_{N/J,a}=v_F+\frac{g_{4,a}(0)\pm g_{2,a}(0)}{2\pi} \; .
\label{TLvelocities}
\end{equation}
The part of  $H^{\rm TL}_a$ involving bosonic operators can be diagonalized by a 
Bogoliubov transformation\cite{Schoenhammer05} leading to a Hamiltonian of free 
and uncoupled bosons with dispersion $\omega_{a}(k)=v_a(k)|k|$ and velocities
\begin{eqnarray}
v_a(k) & & =
v_F \left( 1+\frac{g_{4,a}(k)+g_{2,a}(k)}{2\pi v_F} 
\right)^{1/2} \nonumber \\ 
&& \times \left( 1+\frac{g_{4,a}(k)-g_{2,a}(k)}{2\pi v_F} \right)^{1/2} \; .
\label{vcs}
\end{eqnarray}
In the  limit $k \to 0$ this leads to the charge and spin velocities defined in 
Eq.\ (\ref{defvsc}). 

In the low-energy limit the details of the momentum dependence of the
couplings are generically irrelevant as long as the
$g_{\nu,\parallel/\perp}(k)$ are slowly varying functions close to
$k=0$ (for an exception see Ref.\ \onlinecite{VM3}). Without loss of 
generality we thus assume that 
\begin{equation} 
g_{\nu,\parallel/\perp}(k) = g_{\nu,\parallel/\perp} 
 \, \Theta(k_c-|k|) \; ,
\label{boxpot}
\end{equation}
with a cutoff $k_c = 2 \pi n_c/L \ll k_F$ of the momentum transfer.
Therefore the velocities $v_a$ are independent of $k$ for all $k \leq k_c$.
With this choice a closed expression for the retarded 
one-particle Green function $\tilde {\mathcal G}^R(x,\omega)$ of 
the homogeneous TLM can be given, which enters the
approximate expression for the conductance through the ring used
here. To obtain the Green function we follow the procedure
introduced in Ref.\ \onlinecite{VM4} (see below). 

\subsection{The Luttinger liquid ring with magnetic flux}

We now consider a ring of finite length $L$ described by the TLM that
is pierced by a magnetic flux $\phi$. The term $H_{{\mathcal J}}^c = \pi v_{J,c} {\mathcal J}_c^2/(2
L)$ in the Hamiltonian Eq.\ (\ref{scs}) must then be replaced by\cite{Loss,PeterS,VM2}
\begin{eqnarray}
H_\mathcal{J}^c & = & \frac{v_{J,c} \pi}{2 L}\left({\mathcal J}_c
-\sqrt{2} \, \frac{\phi}{\pi}\right)^2 \nonumber \\
& = & \frac{v_{J,c} \pi}{2 L} \left({\mathcal J}_c^2 - 2 \, \sqrt{2} 
\, {\mathcal J}_c \, \frac{\phi}{\pi} + 2 \, \left[  \frac{\phi}{\pi}\right]^2
\right)
\label{correctv}
\end{eqnarray}
It is important to note that the characteristic velocity appearing in
the flux dependent part of the Hamiltonian is the current velocity
$v_{J,c}$. This can easily be seen as follows. For a system with an equal 
number of left and right moving electrons $\left< {\mathcal J}_c 
\right> = 0$ and the last term in Eq.\ (\ref{correctv}) determines 
the ground state persistent current $I = - dE_0(\phi)/d \phi$, with 
$E_0$ being the ground state energy. It was shown 
numerically\cite{PeterS,VM2} for lattice models that the velocity 
appearing in the
persistent current is the charge current velocity $v_{J,c}$. 
In Ref.\ \onlinecite{Jagla} mostly $v_F$ was used instead of $v_{J,c}$. 
Only for Galilean invariant systems both velocities are 
equal,\cite{Schick} which indicates that the authors of 
Ref.\ \onlinecite{Jagla} had such systems in mind without 
mentioning it explicitly. 
 
We later need to compute the one-particle Green 
function of the isolated LL ring. The term linear in the flux of
Eq.\ (\ref{correctv}) affects this correlation function. 
Using Eqs.\ (\ref{Jdef}) and (\ref{ndef}) the $\phi$-linear term
can be written as 
\begin{eqnarray}
&& - \frac{v_{J,c} \pi}{2 L} \, 2 \, \sqrt{2} \, {\mathcal J}_c \, 
\frac{\phi}{\pi}
= -  \frac{v_{J,c} \phi}{L} 
\nonumber \\ && \times (\mathcal N_{+,\uparrow} + \mathcal N_{+,\downarrow}
-\mathcal N_{-,\uparrow} - \mathcal N_{-,\downarrow}) 
\label{linterm}
\end{eqnarray} 
and the prefactor of the operator part can be understood as an $\alpha$-dependent 
correction of the chemical potential $\delta \mu_\alpha$. It thus appears as a 
phase factor $\exp{(-i \, \delta \mu_\alpha \, t)}$ in the Green function
[see Eq.\ (\ref{wichtig})].  

\subsection{The conductance of the Luttinger liquid ring}

Next, the finite size LL ring is coupled via tunnel barriers with
hopping matrix elements $t'$  to left and 
right leads at positions $x=0$ and $x=L/2$. For simplicity we
assume equal left and right tunnel barriers. To be specific 
both leads are described as 1d tight-binding chains with hopping matrix 
element $\tau$, but we expect our results to be independent of the 
precise form of the band structure of the reservoirs. 

For noninteracting 
electrons the transmission probability per spin direction 
through the ring can be expressed 
in terms of the spin-independent retarded one-particle Green function 
$\tilde {\mathcal G}^{R}(x,\omega)$ of the isolated ring at position $0$ 
and $L/2$. It is given 
by\cite{Jagla} (for a simple derivation of this formula see Sec. III of 
Ref.\ \onlinecite{Tilman})  
\begin{equation}
T(\omega)=\frac{4\tau^2\sin^2k\left|Lt'^2\tilde {\mathcal G}^R(L/2,\omega)
\right|^2}{\left|[\omega-Lt'^2 \tilde {\mathcal G}^R(0,\omega)+\tau e^{ik}]^2-
\left|Lt'^2 \tilde {\mathcal G}^R(L/2,\omega)\right|^2\right|^2} \; .
\label{transmittance1}
\end{equation}
The zero temperature conductance, on which we focus, follows by setting
$\omega=0$, that is the energy equal to the chemical potential, and 
multiplying $T$ by $2 e^2/h$. In Refs.\ \onlinecite{Jagla} and 
\onlinecite{Hallberg} it was argued that this expression can also be 
used in the presence of interactions in the ring provided $t'$ is 
sufficiently small and the Kondo effect does not play a role. The 
latter holds in parameter regimes with an average number of 
electrons on the ring which is even. That the general features of the
conductance through an interacting ring which is not in the Kondo
regime are indeed captured to some extent by the approximate
Eq.~(\ref{transmittance1}) was shown in Ref.\ \onlinecite{Simon}
using a functional renormalization group approach which can be applied
for sufficiently weak interactions but is nonperturbative in the
coupling to the leads. Thus we will here also rely on 
Eq.~(\ref{transmittance1}).

\subsection{The Green function of the $g_4$-model}

In a first step we study the
so-called $g_4$-model in which all inter-branch scattering processes
are set to zero.   
We decompose the retarded Green function as 
\begin{eqnarray}
 i \, {\mathcal G}^R(x,t) = \theta(t) \sum_{\alpha = \pm} 
\left[ i \, {\mathcal G}_\alpha^>(x,t) + i \, {\mathcal G}_\alpha^<(x,t)  
\right] \; ,
\label{decomp}
\end{eqnarray}
with the greater $ {\mathcal G}_\alpha^> $ and lesser 
$ {\mathcal G}_\alpha^<$ Green
functions of right ($\alpha = +$) and left ($\alpha=-$) moving 
electrons. Using bosonization of the field 
operator\cite{Schoenhammer05,Herbert} for the box-shaped 
potential Eq.\ (\ref{boxpot}), these Green functions are 
given by\cite{VM4}
\begin{equation}
i \, \mathcal G_\alpha^{>/<}(x,t)=i \, e^{i \alpha \phi v_{J,c} t/L}  
\mathcal G_\alpha^{>/<,0}(x,t)e^{F^{>/<}_\alpha(x,t)}
\label{wichtig}
\end{equation}
where
\begin{eqnarray}
F^{>/<}_\alpha(x,t)=\sum_{n=1}^{n_c}\frac{1}{n}
\left[\frac{1}{2} e^{  \pm i \,  \frac{2\pi}{L} n (\alpha x- v_c t) }
\nonumber \right. 
\\ \left. + \frac{1}{2} e^{  \pm i\, \frac{2\pi}{L} n (\alpha x- v_s t) } 
- e^{  \pm i\,   \frac{2\pi}{L} n (\alpha x- v_F t)}\right] \; .\label{F(x,t)}
\label{Fxt}
\end{eqnarray}
The noninteracting Green functions read
\begin{eqnarray}
i \, \mathcal G_{\alpha}^{ >,0}(x,t)  & = & 
\frac{1}{L} 
e^{i \alpha k_F  x} e^{ i\, \frac{2\pi}{L}   
 \left( \alpha x - v_F t \right) } \nonumber \\ 
&& 
\times 
\left[ 1-  e^{ i\,  \frac{2\pi}{L}  
 \left( \alpha x - v_F t +i0 \right) } \right]^{-1}
 \; , \nonumber \\
i \, \mathcal G_{\alpha}^{ <,0}(x,t) & = &  
\frac{1}{L}    
e^{i \alpha k_F  x} \nonumber \\
&& \times 
\left[ 1-  e^{ - i\,  \frac{2\pi}{L}  
 \left( \alpha x - v_F t -i0 \right) } \right]^{-1} \; .
\label{gberechnet0}
\end{eqnarray}   
The second exponential factor in the equation for $ \mathcal G_{\alpha}^{ >,0}$
appears due to the choice of $\pm k_F=\pm 2 \pi n_F/L$ corresponding to the 
last occupied level. For 
$L \to \infty$ it becomes unity, but is relevant in our case as we study 
rings of finite length.
As discussed in 
Ref.\ \onlinecite{VM4} the Fourier transform of the greater 
Green function can be computed iteratively as 
\begin{align}
& \tilde {\mathcal G}^>_\alpha(x,\omega)=
\frac{e^{i \alpha (k_F+\frac{2\pi}{L})x}}{L} \sum_{m=0}^\infty\sum_{l=0}^{m}\sum_{j=0}^{m-l}  \nonumber\\
& \frac{a_{m-l-j}^{(n_c)}b_l^{(n_c)}b_j^{(n_c)}e^{i \alpha \frac{2\pi}{L}mx}}{\omega 
+ \alpha \phi v_{J,c}/L 
-\frac{2\pi}{L}\big\lbrack (m-l-j+1)v_F+lv_c+jv_s 
\big\rbrack+i0} \; , \label{erstegreater}
\end{align}
with 
\begin{eqnarray}
a_{lm+n}^{(m)} & = & \sum_{j=0}^l\frac{(-1/m)^j}{j!}a_{m(l-j)+n}^{(m-1)} \; , \nonumber \\
b_{lm+n}^{(m)} & = & \sum_{j=0}^l\frac{(1/2m)^j}{j!}b_{m(l-j)+n}^{(m-1)}
\label{coeffs}
\end{eqnarray}
and initial values
\begin{eqnarray}
a_{m}^{(1)} & = & \sum_{j=0}^m\frac{(-1/j)}{j!} \; , \nonumber\\
b_m^{(1)} & = & \frac{(1/2)^m}{m!} \; .
\end{eqnarray}
Similarly, the lesser Green function reads
\begin{align}
&  \tilde {\mathcal G}^<_\alpha(x,\omega)=
\frac{e^{i \alpha k_F x}}{L} \sum_{m=0}^\infty\sum_{l=0}^{m}\sum_{j=0}^{m-l}  \nonumber\\
& \frac{a_{m-l-j}^{(n_c)}b_l^{(n_c)}b_j^{(n_c)}e^{- i \alpha \frac{2\pi}{L}mx}}{\omega 
+ \alpha \phi v_{J,c}/L 
+ \frac{2\pi}{L}\big\lbrack (m-l-j)v_F+lv_c+jv_s 
\big\rbrack+i0} \; .  \label{erstelesser}
\end{align}

It is easy to see that for $|k\pm k_F| < q_c$ 
the one-particle spectral function as a function of energy obtained 
by taking the imaginary part of the momentum Fourier transform of 
Eqs.\ (\ref{erstegreater}) and (\ref{erstelesser}) has support 
only between $ \pm v_s  |k\pm k_F| $ and $\pm v_c  |k\pm k_F|$.\cite{VM4}   
In the thermodynamic limit the spectral weight shows a square-root 
singularity at the two edges. The corresponding
Green functions (for $|x|, v_F |t| \gg 1/q_c$) in the $(x,t)$-plane are given by 
\begin{eqnarray}
&& i \, {\mathcal G}^{>}_\alpha(x,t)  =  
\frac{1}{L} e^{i \alpha k_F x} e^{i \alpha \phi v_{J,c} t/L}  e^{ i\, \frac{2\pi}{L}   
 \left( \alpha x - v_F t \right) }
\nonumber \\ && 
\times \left[ 1- e^{i  \frac{2 \pi}{L} (\alpha x- v_c t +i 0)}\right]^{-1/2}
 \left[ 1- e^{i \frac{2 \pi}{L} (\alpha x- v_s t +i0)}\right]^{-1/2} 
\; ,\nonumber \\  
&& i \, {\mathcal G}^{<}_\alpha(x,t)  =  
 \frac{1}{L} e^{i \alpha k_F x} e^{i \alpha \phi v_{J,c} t/L}  
 \nonumber \\ && \times
 \left[ 1- e^{-i  \frac{2 \pi}{L} (\alpha x- v_c t -i 0)}\right]^{-1/2}
 \left[ 1- e^{-i \frac{2 \pi}{L} (\alpha x- v_s t -i0)}\right]^{-1/2}
\; . 
\label{terme}
\end{eqnarray}
Power-laws with anomalous (interaction 
dependent) exponents only appear if also 
the $g_2$ terms are kept.\cite{Schoenhammer05} 
Below we return to Eq.\ (\ref{terme}) 
when discussing the results of Ref.\ \onlinecite{Jagla}.
As the anomalous contributions to the propagator were neglected in 
this work, it is the $g_4$-model which was effectively studied.

From Eqs.\ (\ref{TLvelocities}) and (\ref{vcs}) it follows that 
in the $g_4$-model the charge current velocity $v_{J,c}$ and the 
charge velocity $v_c$ are equal. 

\subsection{Green function of the full model}

Also for the full model with intra- and inter-branch 
scattering processes a closed iterative expression for the Green 
function can be given if a box-shaped two-particle interaction
is assumed. Now, the anomalous dimensions $\gamma_s=s_s^2$ and 
$\gamma_c=s_c^2$ appear. The variables $s_a$ are defined by 
$s_a(q)=\sinh\Theta_{q,a}$ at $q=0$ and the angle $\Theta_{q,a}$ 
characterizes via
\begin{equation}
\beta_{q,a} =  b_{q,a}\cosh\Theta_{q,a}-b_{-q,a}^\dagger\sinh\Theta_{q,a}
\end{equation} 
the canonical transformation to new annihilation (creation) operators 
$\beta_{q,a}^{(\dagger)}$ by means of which the bosonized 
TL-Hamiltonian is diagonalized. The angle $\Theta_{q,a}$ 
depends on the couplings according to the formula
\begin{eqnarray}
\tanh(2\Theta_{q,a}) & = & -\Big\lbrack \frac{g_{2,\parallel}(q)\pm g_{2,\perp}(q)}{2\pi v_F} 
\Big\rbrack  \nonumber \\ && \times 
 \Big\lbrack 1+\frac{g_{4,\parallel}(q)\pm g_{4,\perp}(q)}{2\pi v_F} \Big\rbrack^{-1}
\end{eqnarray}
where the ``$+$''-signs are relevant for the charge angle $\Theta_{q,c}$ 
and the ``$-$''signs for the spin angle $\Theta_{q,s}$. The function 
$F^{>/<}_\alpha(x,t)$ in Eq.\ (\ref{wichtig}) now reads
\begin{align}
F^{>/<}_\alpha&(x,t)=\sum_{n=1}^{n_c}\frac{1}{n}\Big\lbrack\frac{1+s_c^2}{2}
 e^{\pm i\frac{2\pi}{L}n(\alpha x-v_ct)}\nonumber\\&+\frac{s_c^2}{2}
 e^{\mp i\frac{2\pi}{L}n(\alpha x+v_ct)}
+\frac{1+s_s^2}{2} e^{\pm i\frac{2\pi}{L}n(\alpha x-v_st)}
\label{Fneu}\\
&+\frac{s_s^2}{2} e^{\mp i\frac{2\pi}{L}n(\alpha x+v_st)}
-e^{\pm i\frac{2\pi}{L}n(\alpha x-v_Ft)}-s_c^2-s_s^2\Big\rbrack.\nonumber
\end{align}
Because of the assumed box-like shape of the two-particle interaction (in momentum space) 
the $s_a(q)$ are independent of $q=q_n=2 \pi n/L$ for all $n \leq n_c$. 

We again introduce recursively defined coefficients. The $a_m^{(n_c)}$ are given
as in  Eq.\ (\ref{coeffs}) and coefficients $b_{m,a}^{(n_c)}$ 
as well as $c_{m,a}^{(n_c)}$ are computed according to the formulas
\begin{equation}
\begin{split}
b_{lm+i,a}^{(m)}=\sum_{j=0}^l\frac{\left( [1+\gamma_a]/[2m] \right)^j}{j!}b_{m(l-j)+i,a}^{(m-1)}\;,\\
c_{lm+i,a}^{(m)}=\sum_{j=0}^l\frac{\left( \gamma_a/[2m] \right)^j}{j!}c_{m(l-j)+i,a}^{(m-1)}
\end{split}
\end{equation}
with initial conditions
\begin{equation}
\begin{split}
b_{m,a}^{(1)}=\frac{\left( [1+\gamma_a]/2 \right)^m}{m!},\\
c_{m,a}^{(1)}=\frac{\left( \gamma_a/2 \right)^m}{m!}.
\end{split}
\end{equation}
In analogy to Eq.\ (\ref{erstegreater}) 
one can bring the greater and lesser Green functions into the form 
\begin{equation}
\begin{split}
&\tilde{{\mathcal G}}^>_\alpha(x,\omega)=A^{-\gamma_c-\gamma_s}\frac{e^{i\alpha(k_F+\frac{2\pi}{L})x}}{L}\\
&\sum_{m=0}^\infty\sum_{l=0}^{m}\sum_{j=0}^{l}\sum_{p=0}^{m-l}\sum_{q=0}^{p}
a_{m-l-p}^{(n_c)}b_{l-j,c}^{(n_c)}c_{j,c}^{(n_c)}b_{p-q,s}^{(n_c)}c_{q,s}^{(n_c)}\\
&\frac{e^{i \alpha \frac{2\pi}{L} (m-2j-2q)x}}{\omega+ \alpha \phi v_{J,c}/L 
-\frac{2\pi}{L}\left( \big\lbrack(m+1-l-p)v_F-lv_c-pv_s\big\rbrack \right)+i0}\label{zweitegreater}
\end{split}
\end{equation}
and
\begin{equation}
\begin{split}
&\tilde{\mathcal{G}}^<_\alpha(x,\omega)=A^{-\gamma_c-\gamma_s}\frac{e^{i\alpha k_Fx}}{L}\\
&\sum_{m=0}^\infty\sum_{l=0}^{m}\sum_{j=0}^{l}\sum_{p=0}^{m-l}\sum_{q=0}^{p}
a_{m-l-p}^{(n_c)}b_{l-j,c}^{(n_c)}c_{j,c}^{(n_c)}b_{p-q,s}^{(n_c)}c_{q,s}^{(n_c)}\\
&\frac{e^{-i \alpha \frac{2\pi}{L} (m-2j-2q)x}}{\omega+ \alpha \phi v_{J,c}/L 
+\frac{2\pi}{L}\left( \big\lbrack(m-l-p)v_F-lv_c-pv_s\big\rbrack \right)+i0}\label{zweitelesser}
\end{split}
\end{equation}

\section{The approximation of Jagla and Balseiro}
\label{jaglaapprox}

As mentioned in Sec.\ \ref{model}, in their work Jagla and Balseiro (JB)\cite{Jagla} 
effectively studied the $g_4$-model as they neglected the 
anomalous contributions 
to the propagator. The starting point is the retarded Green function given by the
sum of the approximate expressions (\ref{terme}) (valid for $|x|, v_F |t| \gg 1/q_c$), 
where in addition $v_{J,c}$ is replaced by $v_F$. These two velocities 
become equal in Galilean invariant systems with $g_{4,c}(0)=g_{2,c}(0)$ 
(and trivially in the noninteracting case). 
This condition cannot be achieved within the $g_4$-model for any reasonable 
choice of $g_{4,\parallel}$ and $g_{4,\perp}$. It 
is thus to some extend inconsistent to replace the charge current velocity by 
the noninteracting Fermi velocity. 
Compared to Eq.\ (\ref{terme}), JB also seem to have
neglected the factor $ \exp{ \left\{ i\, \frac{2\pi}{L}   
 \left( \alpha x - v_F t \right) \right\} } $ appearing 
in $\mathcal G_{\alpha}^{ >}(x,t)$ [see Eq.\ (6) of Ref.\ \onlinecite{Jagla}] which
has to be included due to the finiteness of the system.

To analytically perform the Fourier transformation
of the approximate Green function from time to frequency 
JB only consider cases where
$v_c/v_s=q/p$, with $q$ and $p$ being ``small'' odd numbers, which corresponds to 
rather specific choices of two-particle couplings. 
They {\it assume} that the Fourier transform is essentially determined by the 
behavior of the ``dominant poles''. For $x=L/2$ poles of the form
$1/(t-t^\ast)$ appear at $t^\ast=qL/(2 v_c) = p L/(2 v_s)$, that
is, times at which charge and spin excitations starting (at the same time) from the 
left contact and traveling along the ring with $v_c$ and $v_s$ get
together at the right contact. One expects this approximation
to work best for small numbers $p$ and $q$ because otherwise such poles are
rare among all poles of $\mathcal G_{\alpha}^{ >}(x,t)$. The function
\begin{align}
& \frac{\Theta(t)}{T_1 \sqrt{v_c v_s}}\sum_{\alpha=\pm1} 
e^{i\alpha k_F x}e^{i\alpha\frac{\phi v_{J,c}}{L} t } \nonumber \\
& \times \left[ \frac{e^{i \frac{2\pi}{L}(\alpha x-v_Ft)}}
{1-e^{\frac{2\pi i}{L}(\alpha x- p v_ct+i0)}}+\frac{1}
{1-e^{-\frac{2\pi i}{L}(\alpha x-pv_ct-i0)}}\right]
\label{polpfusch} 
\end{align}
has for $x=0$ and $x=\frac{L}{2}$ the same ``dominant poles'' $t^\ast$ 
and residua as 
the Green function resulting from summing up the two terms of 
Eq.\ (\ref{terme}). The time between two such poles is given 
by $T_1=\frac{qv_F}{v_c}T_0=\frac{pv_F}{v_s}T_0$. The 
time $T_0=v_F/L$ is the $\phi=0$ period (in $t$) of the Green function in the 
noninteracting case, while $T_1$ is the corresponding period of the 
approximate expression (\ref{polpfusch}) including the interaction. 
Fourier transforming Eq.\ (\ref{polpfusch}) leads to 
\begin{eqnarray}
& \tilde {\mathcal G}^R(x,\omega) & = 
\frac{1}{T_1\sqrt{v_cv_s}} \sum_{\alpha=\pm 1} \left(e^{i\alpha k_Fx}  \right.\nonumber\\
&&\left.\sum_{m=0}^\infty \left[ 
\frac{e^{i\alpha \frac{2\pi}{L}(m+1)x}}{\omega+
\alpha\phi v_{J,c}/L - \frac{2\pi}{L}(m+1)\frac{L}{T_1}+i0} \right. \right. \nonumber\\
&&+\left. \left. \frac{e^{-i\alpha \frac{2\pi}{L}mx}}{\omega+
\alpha\phi v_{J,c}/L + \frac{2\pi}{L}m\frac{L}{T_1}+i0}\right]\right) \; .
\label{polpfusch2}
\end{eqnarray}

Within this ``dominant pole'' approximation {\it and} for $v_{J,c} \to v_F$, 
the shape of the transmission probability Eq.\ (\ref{transmittance1}) depends 
almost exclusively on $T_1/T_0$. Replacing $T_0$ by $T_1$, mainly rescales 
the $\phi$-axis by a factor $T_0/T_1$. In the noninteracting case 
$T(\phi)$ is a periodic function of period $2 \pi$ which is symmetric 
with respect to $\phi=\pi$. 
Replacing $\phi$ by $\phi+2\pi m \,  T_0/T_1$ 
so that the new magnetic flux is still within $[0,2\pi]$ the transmission probability remains 
invariant as one has effectively shifted the summation variable  
$m \to m+n$ in Eq.\ (\ref{polpfusch2}).
Thus, the sole effect of the different interaction dependent velocities 
is to reduce the periodicity of the magnetic flux dependence of the 
conductance by a factor $T_0/T_1$. Since in the noninteracting case the 
number of dips of $G(\phi) =e^2 T(\phi)/h$ in the interval of periodicity equals 
$1$ (see Fig.\,\ref{noint}), it becomes $T_1/T_0$ in the presence of interactions according 
to the ``dominant pole'' approximation. This effect is shown in Figs.\ 2 and 3 
of Ref.\ \onlinecite{Jagla}.

To some extend the ``dominant pole'' approximation is motivated by the simple
physical picture that the spin (traveling with $v_s$) and the charge excitations 
(traveling with $v_c$) have to ``recombine'' at $x=L/2$ (and $x=0$) for an electron to pass 
the LL ring. Indeed, the poles kept in the present 
approximations correspond to such times. We next show that the usefulness of this 
picture is quite limited and that the number of dips of the conductance resulting from an exact 
evaluation of the Green function differs considerably from the above result  
even in the $g_4$-model.

\section{Results}
\label{results}

We now discuss the conductance (transmission) through the ring described 
by the TLM and coupled to leads at positions $x=0$ and $x=L/2$. We use 
the approximate expression (\ref{transmittance1}) relating the retarded 
Green function and the transmission probability. 

The transmission probability is averaged over a small energy window 
around $\omega=0$. This has also been done 
in Refs.\ \onlinecite{Jagla} and \onlinecite{Hallberg}. 
Hallberg {\it et al.}\cite{Hallberg} argue that such an averaging 
over a finite energy window accounts for ``possible (gate and bias) 
voltage fluctuations and temperature effects unavoidably present in an experimental 
system''. 

Our energy unit is fixed by setting the hopping in the leads to 
$\tau=1$.
Transmission curves are only weakly dependent on the length of the ring 
$L$, provided $L$ is not too small. Calculations in this paper are performed for 
$L=256 v_F$ which turns out to be sufficiently large to observe structures as described 
in Ref.\ \onlinecite{Jagla}. To bring out well discernible dips in the transmission probability
and thus results comparable to those obtained by Jagla and Balseiro, 
who neither give their $t'$ nor the size of the energy window averaged over, 
we choose $t'^2=0.005$ and take the average over the 
$\omega$-interval $[-\omega_{\rm{max}},\;\omega_{\rm{max}}]=[-0.05,\;0.05]$ 
if not mentioned otherwise. The width of the box potential
is determined by $n_c$, see Eq.~(\ref{boxpot}), which we choose 
to be $n_c=5$ but since we are dealing only with low-energy properties 
of the system, all results are practically independent of the width 
of the box potential, as we have also checked numerically.

\begin{figure}[hbt]
\begin{center}
\begin{minipage}{8cm}
 \epsfig{file=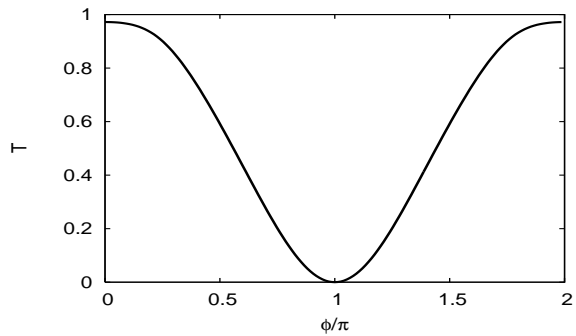,width=8cm,height=4.5cm}
\end{minipage}
\end{center}
\caption{Transmission probability for the noninteracting case in which all characteristic 
velocities become equal to $v_F$.}\label{noint}
\end{figure}

Figure \ref{noint} shows the transmission probability as a function of the magnetic flux
in the noninteracting case, i.\,e., in the case where all characteristic velocities
are equal to $v_F$. It displays a ``dip'' 
at $\phi=\pi$ which occurs independently of the 
chosen coupling to the leads
and the averaging frequency interval $[-\omega_{\rm{max}},\;\omega_{\rm{max}}]$. 
Varying these parameters either renormalizes the curve as a whole
or makes the dip sharper or wider.

\subsection{Results for the $g_4$-model}

We next compute the transmission  probability Eq.\ (\ref{transmittance1}) as a function of $\phi$ 
using the {\it exact} Green function [see Eqs.\ (\ref{decomp}), (\ref{erstegreater}), 
and (\ref{erstelesser})] of the $g_4$-model with a box-shaped two-particle potential.

\begin{figure}[hbt]
\begin{center}
\begin{minipage}{8cm}
 \epsfig{file=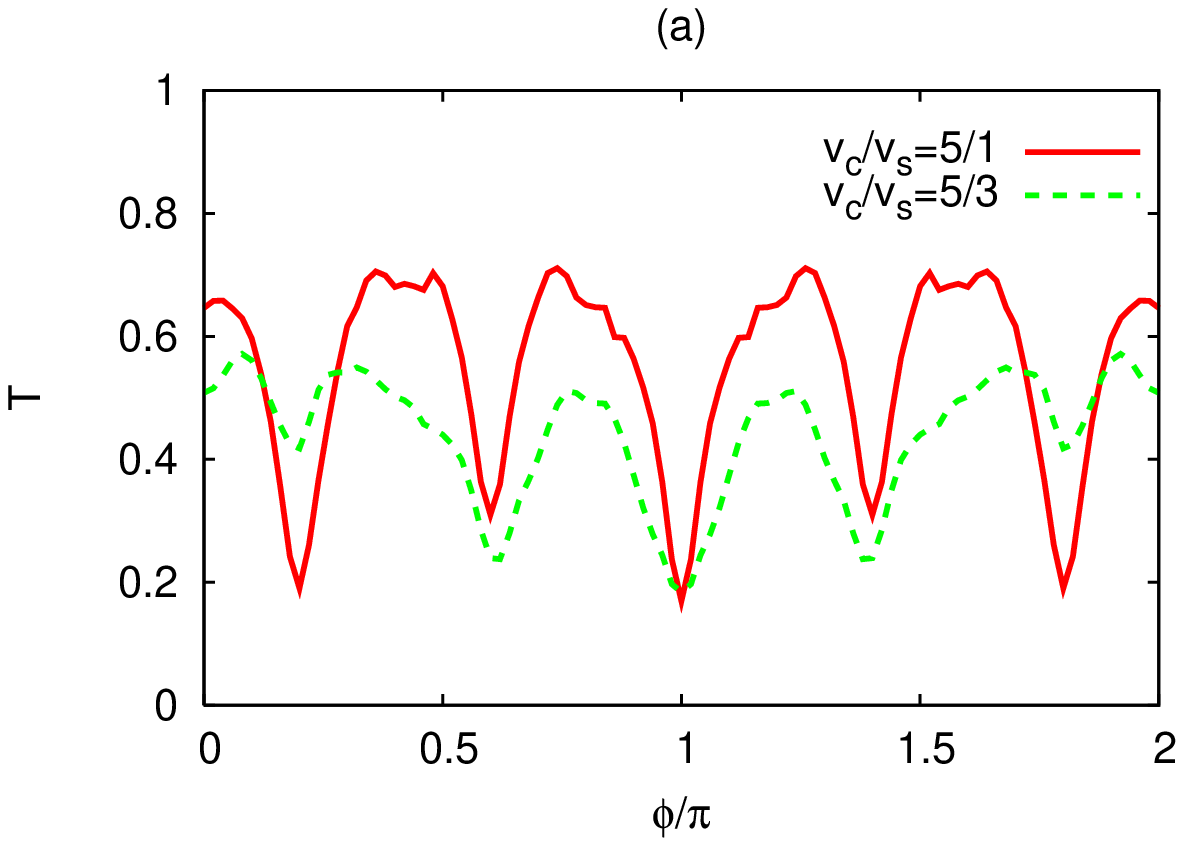,width=8cm,height=3.5cm}
\end{minipage}
\begin{minipage}{8cm}
 \epsfig{file=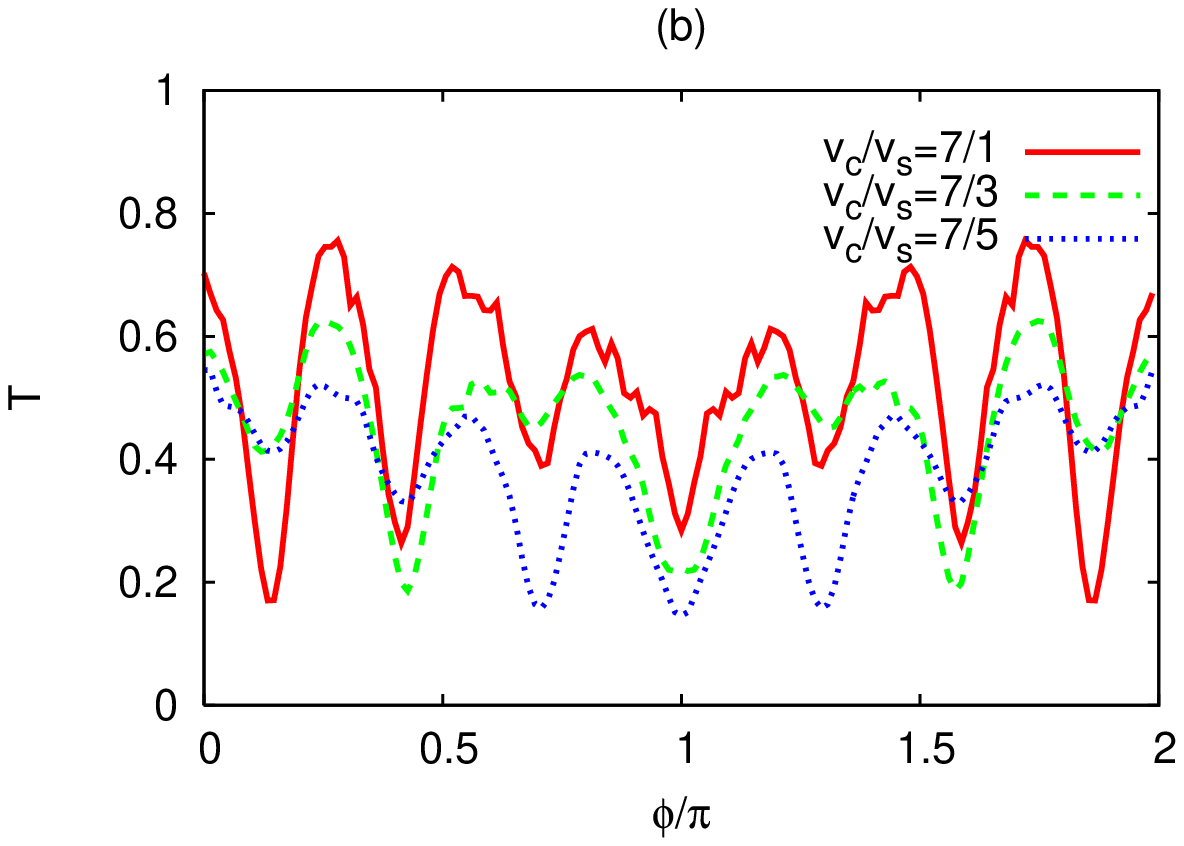,width=8cm,height=3.5cm}
\end{minipage}
\begin{minipage}{8cm}
 \epsfig{file=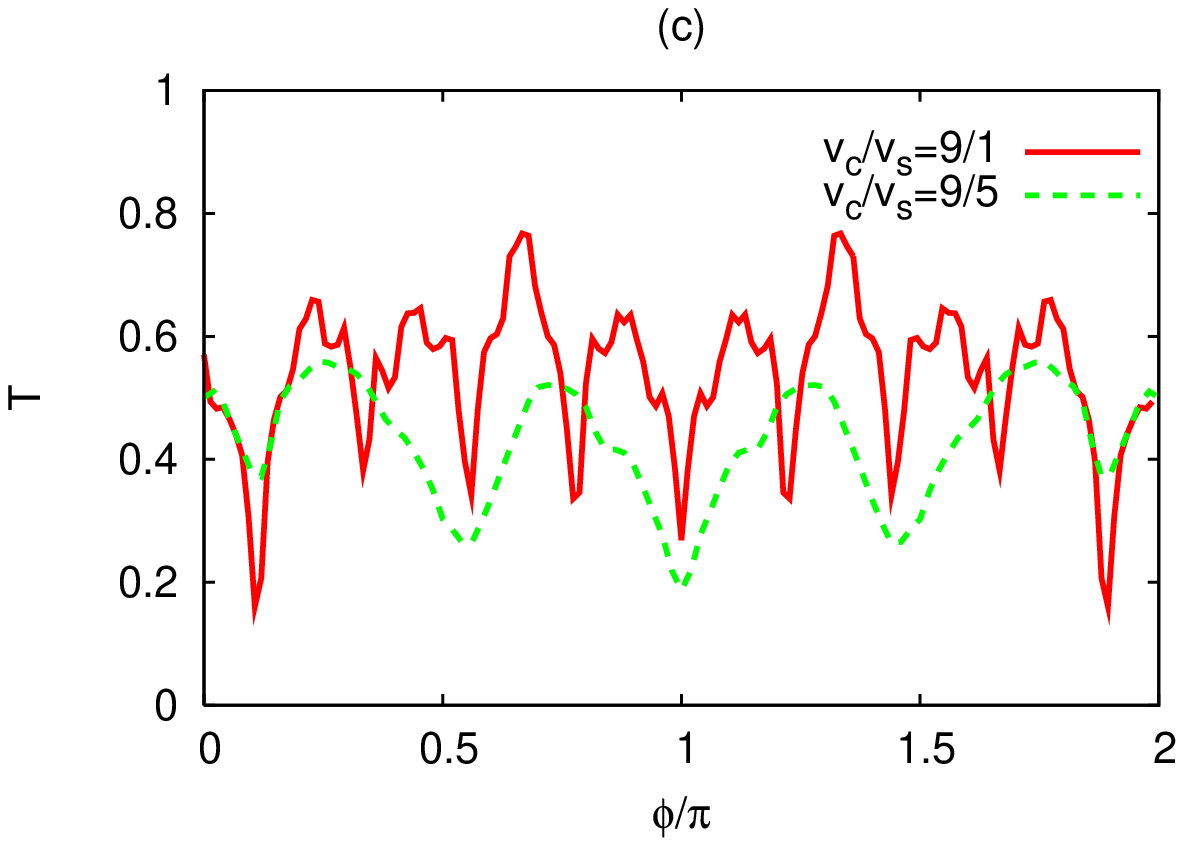,width=8cm,height=3.5cm}
\end{minipage}
\end{center}
\caption{ (Color online) 
Transmission probability for the $g_4$-model with $v_{J,c}/v_s=v_c/v_s=q/p$, $v_{J,c}/v_F=1$ 
and $T_1/T_0=qv_F/v_c=p v_F/v_s=5,\-7,\-9$ from top to bottom. The number of prominent 
dips is always equal to $T_1/T_0$ with the exception of the case where $v_c/v_s=9/5$ 
(see the text). The coupling to the leads is $t'^2=0.001$ in these pictures to minimize
the dependence of the curves on the energy interval over which the 
average is taken.}\label{polok}
\end{figure}

We first consider cases where the interaction parameters are chosen such that 
$v_c/v_s = q/p$ with  odd integers $q$ and $p$, as well as $v_{J,c}/v_F=1$, i.\,e. cases 
where the point made about which velocity to use when taking into account the magnetic 
flux is irrelevant. This corresponds to the situation mainly studied by Jagla and Balseiro. 
We again emphasize that $v_{J,c}=v_c$ for the $g_4$-model and the charge velocity and the charge 
current velocity can be identified throughout this subsection. 
Our calculations  show that under the above conditions the ``dominant pole'' approximation 
remains a good guide only for $T_1/T_0 \lessapprox 7$ but that it loses its validity for 
$T_1/T_0 \gtrapprox 9$. In particular, for sufficiently small $T_1/T_0$ the number of prominent 
dips within $[0,2 \pi]$ is given by $T_1/T_0=qv_F/v_c=p v_F/v_s$. 
This is shown in Fig.\,\ref{polok} where curves for different values
of $T_1/T_0$ are presented. While for all cases where $T_1/T_0=5$ 
[Fig.\ \ref{polok} (a)] or $T_1/T_0=7$ [Fig.\ \ref{polok} (b)] the number 
of prominent dips does indeed equal $5$ or $7$, respectively, it is $9$ for $T_1/T_0=9$ only if 
$p=1$, i.\,e. $v_c/v_s=9$ [solid line in Fig.\ \ref{polok} (c)]. The example 
of the curve with $v_c/v_s=9/5$ [dashed line in Fig.\ \ref{polok} (c)] shows 
that in this case the numbers $p=5$ and $q=9$ are apparently not small enough
but that the transmission probability rather resembles that of the case where 
$v_c/v_s=5/3$ [dashed line in Fig.\ \ref{polok} (a)]. Further down we  
argue that the latter is not accidental. Our analysis specifies how small 
the odd integers $q$ and $p$ have to be for the  ``dominant pole'' 
approximation  to be applicable. Specifically, we find 
that this approximation  leads to qualitatively 
correct results only if the smaller of the two numbers, say $p$, is either 
$1$, $3$ or $5$, with the refinement that for $p=3$ or $5$
the other number $q$ may not exceed $7$.

\begin{figure}[hbt]
\begin{center}
\begin{minipage}{8cm}
 \epsfig{file=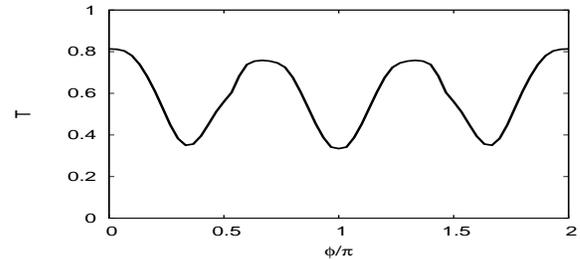,width=8cm,height=3.5cm}
\end{minipage}
\end{center}
\caption{ Transmission probability for the $g_4$-model with $v_c/v_s=3/1$, $v_{J,c}/v_F=v_c/v_F=3/2>1$ 
and  $(v_{J,c}/v_F)(T_1/T_0)=3$ dips, while $T_1/T_0=2$.}\label{neufig}
\end{figure}

We now proceed and consider situations in which the coupling constants are still chosen 
such that  $v_c/v_s=q/p$, but $v_{J,c}/v_F = v_c/v_F > 1$ (for the $g_4$-model with repulsive 
two-particle interaction). The factor $v_{J,c}$ in front of the magnetic flux in Eq.\ (\ref{polpfusch2}), 
leads to a decrease of the periodicity of $T(\phi)$ by a factor $v_F/v_{J,c}$ and thus to 
an increase of the number of dips by a factor $v_{J,c}/v_F$  with respect to $T_1/T_0$, the result 
obtained for $v_F=v_{J,c}$. For sufficiently small $q$ and $p$, that is if the 
``dominant pole'' approximation is applicable, we thus expect to find $(v_{J,c}/v_F)(T_1/T_0)$ 
dips. This is illustrated in Fig.\ \ref{neufig} for $v_{J,c}=v_c = 3v_F/2 = 3v_s$, that is $p=1$, $q=3$.

If the ``dominant pole'' approximation were  applicable for all $p,q$, 
one would thus have $(v_{J,c}/v_F)(T_1/T_0)$ dips in $T(\phi)$ in the interval $\phi \in [0,2\pi]$.
Given any arbitrary combination of the (relevant) velocities $v_s$, $v_c$, $v_{J,c}$, a natural
strategy for guessing the number of dips would be to look for ``small'' odd numbers 
$p$ and $q$ for which $pv_c\approx qv_s$ and to compute  $(v_{J,c}/v_F)(T_1/T_0)$ 
from these. The integers $p$ and $q$ would probably be looked for so that 
$pv_c\approx qv_s$ holds as accurately as possible for, at the same time, 
$p$ and $q$ as small as possible. From the dashed line in Fig.\,\ref{polok} (c) it is 
clear that already the numbers $p=5$ and $q=9$ are not small enough for the ``dominant pole'' 
approximation to be reliable. In such cases, one has to resort to smaller odd 
numbers $p$ and $q$---for which $pv_c\approx qv_s$  holds less well---to predict the number of dips. 
As already shown in Fig.\,\ref{polok} (c), for $v_c=v_F$ and $v_c/v_s=9/5$, instead of $p=5$ and $q=9$ one has to take
$p=3$ and $q=5$ to obtain the correct number of dips. As we shall further show,
it is mostly necessary to choose $p$ and $q$ such that for the smaller 
number $p=1$ even if the approximate identity $pv_c\approx qv_s$ does then hold only
to a very low degree of accuracy. This leads to our central result
\begin{eqnarray}
\label{upshot}
\left. \frac{v_{J,c} T_1}{v_F T_0} \right|_{p=1}= \frac{v_{J,c} v_F}{v_F v_s} = \frac{v_{J,c}}{v_s}
\end{eqnarray}
for the number of dips.

\begin{figure}[hbt]
\begin{center}
\begin{minipage}{8cm}
 \epsfig{file=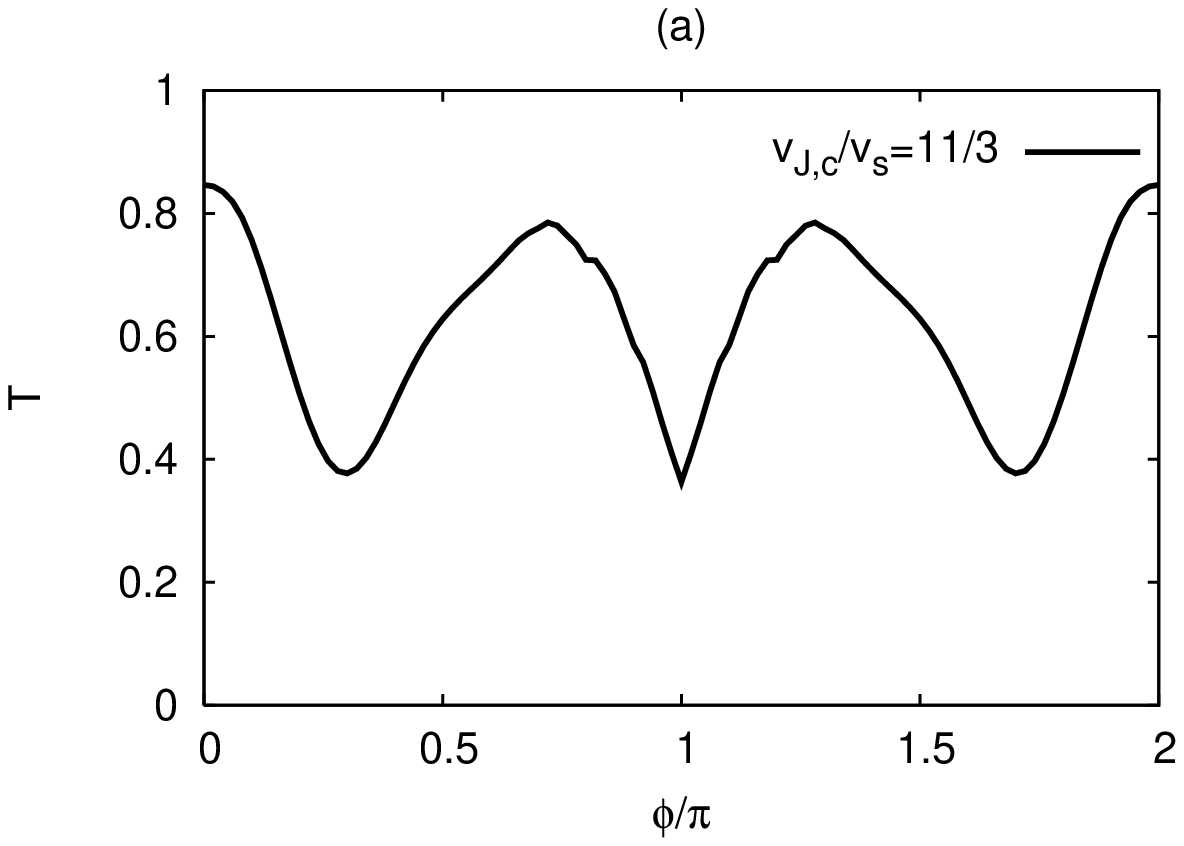,width=8cm,height=3.5cm}
\end{minipage}
\begin{minipage}{8cm}
 \epsfig{file=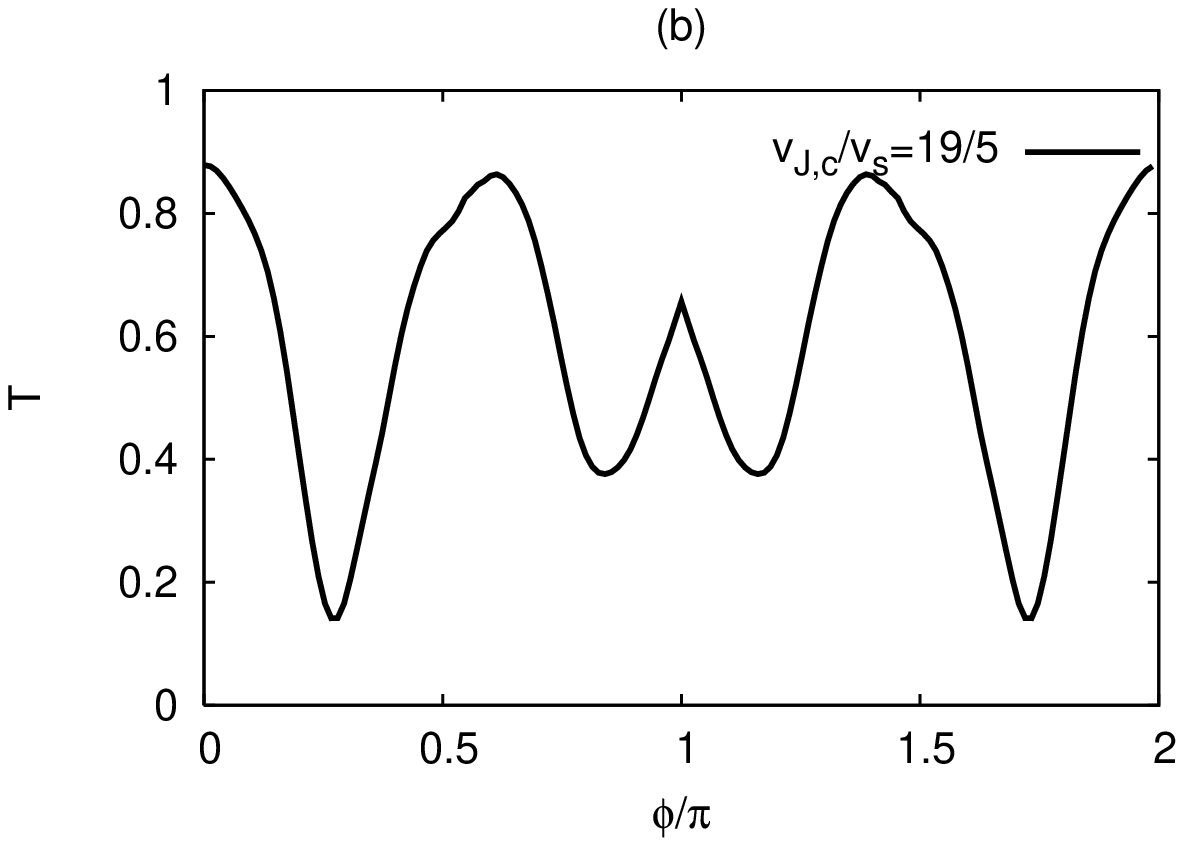,width=8cm,height=3.5cm}
\end{minipage}
\begin{minipage}{8cm}
 \epsfig{file=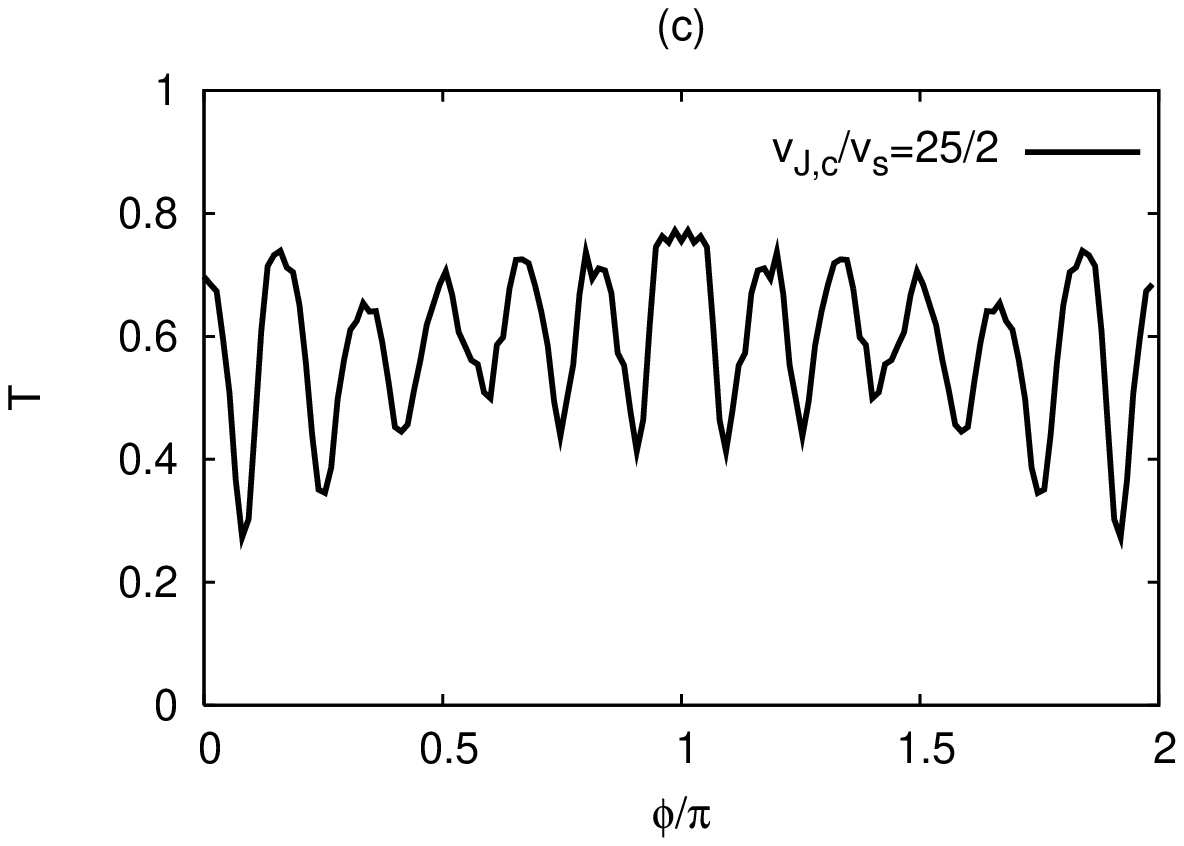,width=8cm,height=3.5cm}
\end{minipage}
\begin{minipage}{8cm}
 \epsfig{file=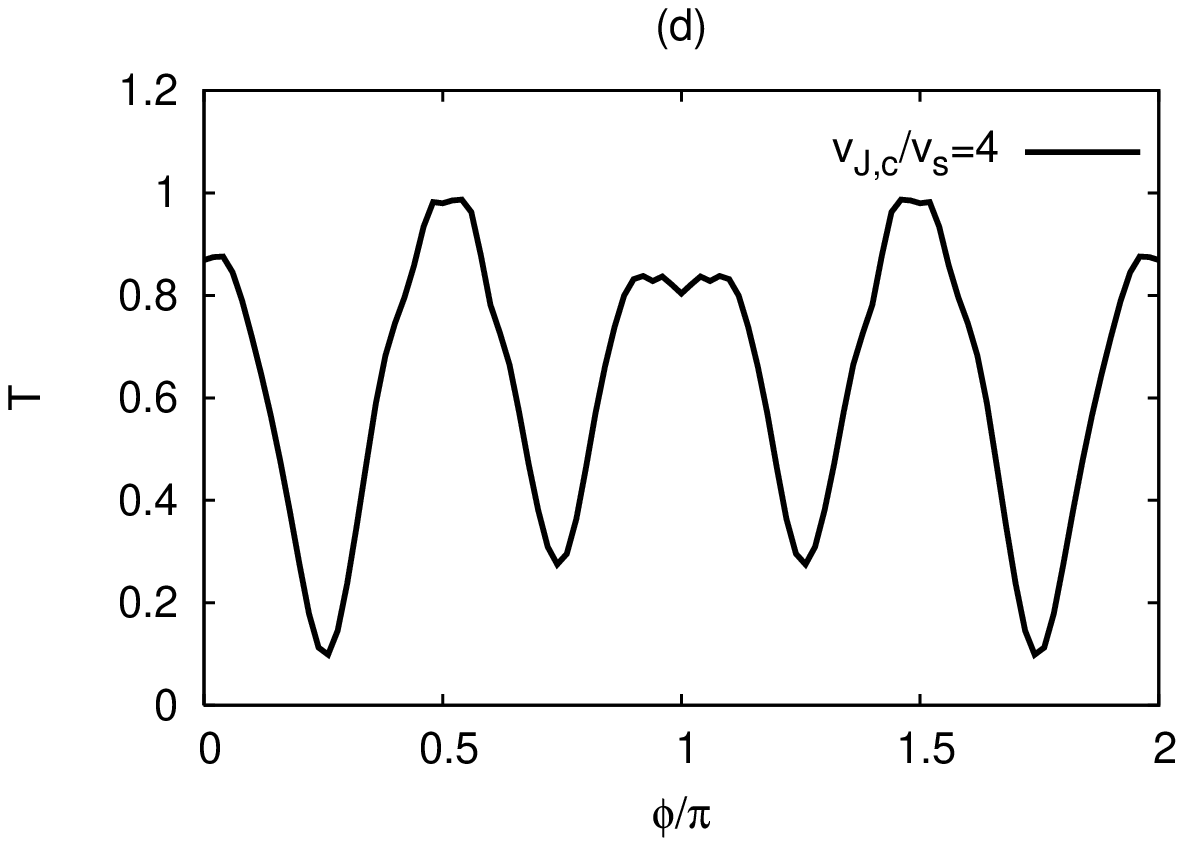,width=8cm,height=3.5cm}
\end{minipage}
\end{center}
\caption{ Transmission probability  for the $g_4$-model with the ratio of charge and spin
velocities as indicated. In all cases, the number of prominent dips is given by one of the 
integers closest to $v_{J,c}/v_s$ ($=v_c/v_s$ for the $g_4$-model).}\label{g4}
\end{figure}

Consider, as an example for the validity of Eq.\ (\ref{upshot}), the case 
where $v_c/v_F=11v_s/(3 v_F) = 11/6$ ($p=3$, $q=11$). We then have
$T_1/T_0=pv_F/v_s=3\cdot2=6$. Taking into account that $v_{J,c}/v_F=v_c/v_F=11/6$, following
the ``dominant pole'' approximation one
would arrive at the conclusion that the number of dips should equal 
$v_{J,c}T_1/(v_F T_0)=11/6\cdot6=11$. In fact, the numbers $p=3$ and $q=11$
are already too large for this approximation to be valuable. Smaller
numbers, for which the equation $pv_c= qv_s$ is still ``approximately'' fulfilled,
are $p=1$ and $q=3$. This gives $\left. T_1/T_0\right|_{p=1}=\left. p v_F/v_s\right|_{p=1} =2$ and thus 
$\left.(v_{J,c}/v_F)(T_1/T_0) \right|_{p=1} =11/3$. The number of dips should, of course,
be an integer number, and the result $11/3$ lies between $3$ and $4$ so that one
can expect one of these two to be the right answer. Indeed, as shown in 
Fig.\,\ref{g4} (a), the corresponding curve exhibits $3$ dips.
This kind of reasoning is applicable whenever the approximate identity $pv_c\approx qv_s$ 
does not hold to some sufficiently high degree of accuracy with
$p=3,\-5$ and $q=5,\-7$ [as it was the case for the example of the dashed line in Fig.\,\ref{polok} (c)]. 
We find that Eq.\ (\ref{upshot}) for estimating the number of dips of $T(\phi)$ 
can be used in almost all cases. Further examples of its validity
are shown in Figs.\,\ref{g4} (b), (c) and (d). For Fig.\ \ref{g4} (b)
$v_{J,c}/v_s=3.8$ and the number of dips is four, while for Fig.\ \ref{g4} (c)
$v_c/v_s=25/2$ so that the numbers $p$ and $q$ for which $pv_c=qv_s$ are no longer both
odd. Equation (\ref{upshot}), however, applies nevertheless and, in this case, leads 
to the result $v_{J,c}/v_s=12.5$ again giving a correct estimate of the number of dips.

Jagla and Balseiro argue that, according to the assumptions (``recombination'' of spin and charge) 
underlying their ``dominant pole'' approximation, the transmission probability would have 
to be ``very small'' 
whenever the couplings are chosen such that $(v_c/v_s)^{\pm 1}=2n$ with integer $n$. In 
these cases
no poles of the form $1/(t-t^\ast)$ appear in $\mathcal G(x=L/2,t)$ and charge and spin
excitation cannot ``recombine'' at finite times at $x=L/2$.\cite{Jagla}
However, the expectation that the transmission probability is small under these conditions
is not confirmed by our calculations. Fig.\ \ref{g4} (d) shows the example of a 
curve for which $v_c/v_s=4/1$ 
and in this case (and in all the others we have checked) the transmission 
probability is not particularly small.
This observation provides further evidence that the idea that charge and spin excitations 
have to ``recombine'' 
at $x=L/2$ for an electron to pass through the ring is of very limited usefulness in the present context.

\subsection{Results for the full model}

\begin{figure}[hbt]
\begin{center}
\begin{minipage}{9cm}
 \epsfig{file=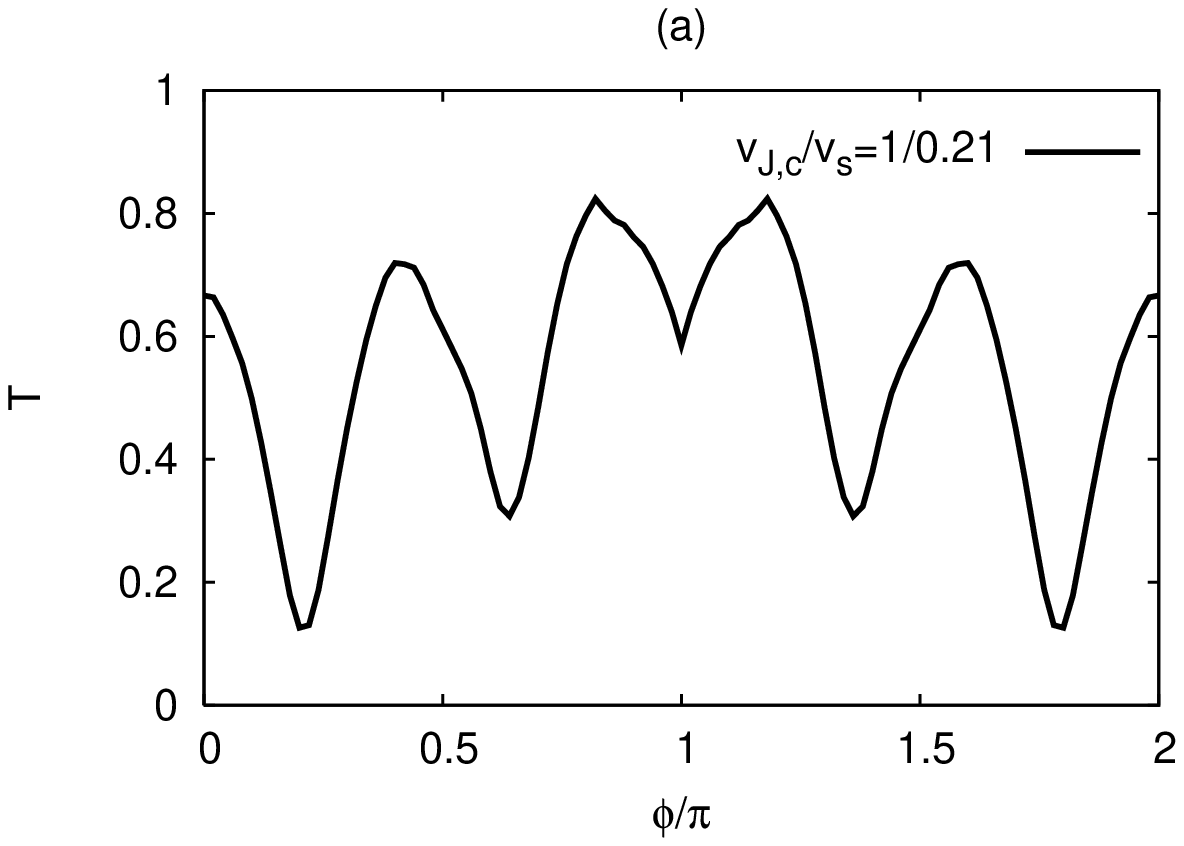,width=9cm,height=3.5cm}
\end{minipage}
\begin{minipage}{9cm}
 \epsfig{file=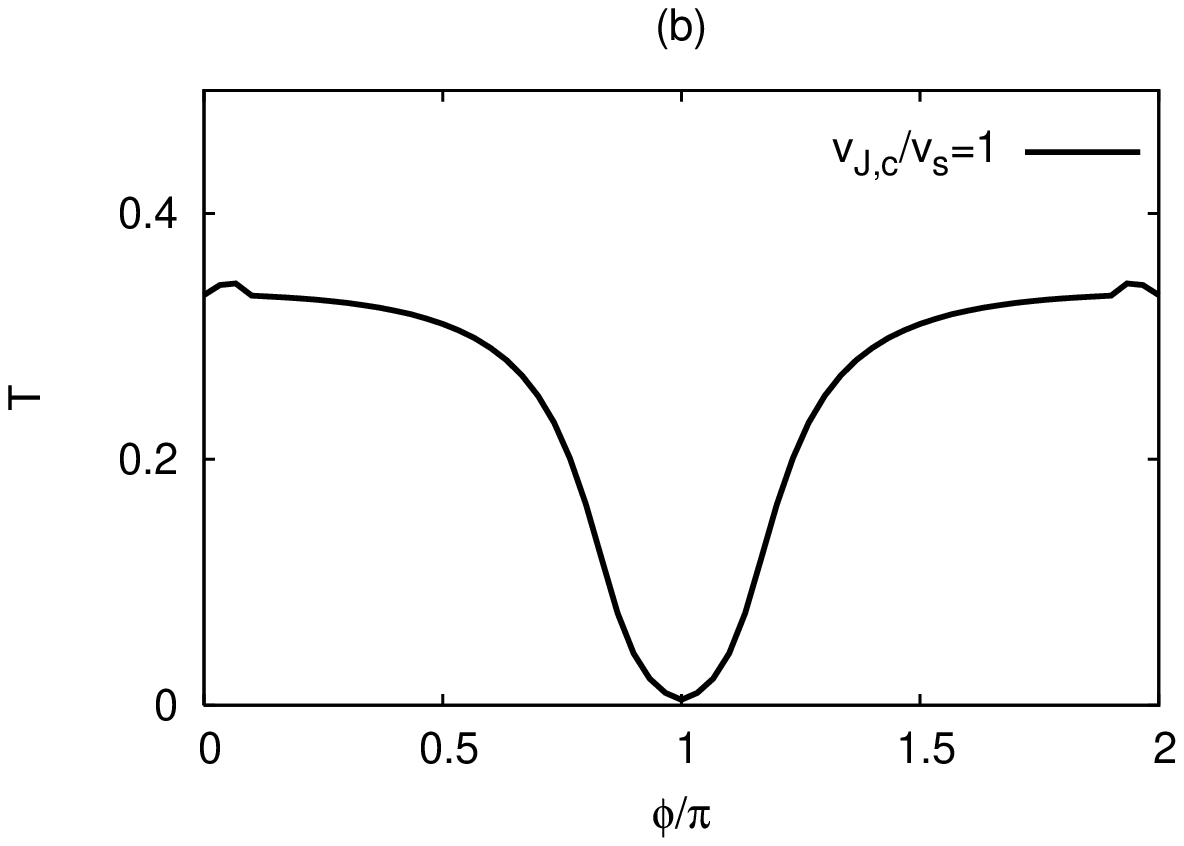,width=9cm,height=3.5cm}
\end{minipage}
\end{center}
\caption{ Transmission probability  for the full model and different sets of coupling parameters.
The individual couplings are chosen such that the (relevant) velocities are $v_{J,c}/v_F=1$, $v_c/v_F \approx  
1.79$, $v_s/v_F  \approx 0.21$ in (a) and  $v_{J,c}/v_F=v_s/v_F=1$, $v_c/v_F \approx 2.01$ in (b).
}\label{g2a}
\end{figure}

All the results obtained in the model with only $g_4$-couplings essentially carry
over to the full model (with additional inter-branch interaction $g_2$), 
the main difference being that now $v_{J,c}\neq v_c$. The charge current velocity $v_{J,c}$ 
Eq.\ (\ref{TLvelocities}) becomes smaller than the charge velocity $v_c$ due to the presence 
of nonvanishing $g_2$-couplings and it is even reduced to the Fermi
velocity if $g_{2,\parallel/\perp}(0)=g_{4,\parallel/\perp}(0)$. 
An example of such a case is shown in Fig.\ \ref{g2a} (a) and the number of dips agrees 
with the prediction made from Eq.~(\ref{upshot}) $v_{J,c}/ v_s = 1/0.21 \approx 5$.

In case \textsl{all} four couplings $g_{\nu,\parallel/\perp}$ are chosen
to be equal, also the spin velocity becomes equal to the Fermi velocity, one
then has $v_s=v_{J,c}=v_F$. The transmission probability for such a case is presented in
Fig.\,\ref{g2a} (b). It shows a dip structure similar to that of the 
noninteracting case. This, however, does not come as a surprise in view of 
the foregoing considerations but is in accordance with Eq.~(\ref{upshot}) 
$v_{J,c}/ v_s = 1$.

\begin{figure}[hbt]
\begin{center}
\begin{minipage}{8cm}
 \epsfig{file=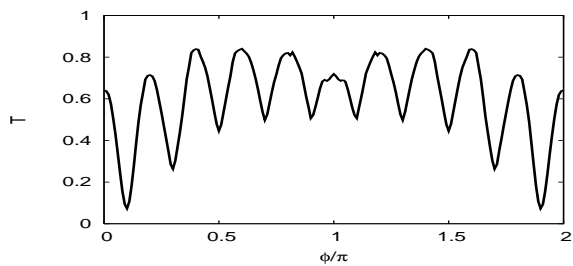,width=8cm,height=3.5cm}
\end{minipage}
\end{center}
\caption{Transmission probability for couplings such that the (relevant) velocities
 are $v_{J,c}/v_F \approx 1.95$, $v_s/v_F \approx 0.195$, and $v_c/v_F \approx 2.32$ leading 
to $v_{J,c}/v_s \approx 1.95/0.195=10$.
}\label{g2b}
\end{figure}

Finally, the transmission probability for a more general case is presented in 
Fig.\,\ref{g2b}. Here, $T(\phi)$ exhibits $10$ dips
and this does indeed represent the ratio of the charge current and the spin
velocities following from the chosen coupling constants, namely
$v_{J,c}/v_s \approx 1.95/0.195=10$. 

Our general finding that in most cases $v_{J,c}/v_s$ gives an estimate of 
the number of dips, may be compared to the result of Hallberg {\it et al.}\cite{Hallberg}. They 
studied the 1d $t$-$J$ model. The Green function of a few lattice site (up to 8 sites) 
isolated $t$-$J$ ring was computed using exact diagonalization. For $J=0$ this model 
corresponds to the $U\rightarrow\infty$\,-\,limit of a one-dimensional Hubbard-Hamiltonian 
for the ring. Whenever the intermediate state contains $N$ electrons, 
the  transmission probability shows $N$ equidistant dips. The authors argue 
that in the limit $J=0$ the ratio of the spin and charge 
velocities becomes $v_c/v_s=N$ and this way try to relate their findings to the ones of 
Ref.~\onlinecite{Jagla}. It remains, however, unclear how this would explain the observed 
dip structure in view of the results presented here which unambiguously suggest 
that $v_{J,c}$ and not $v_c$ is the velocity of importance for the number of dips. In particular, 
in the $U\rightarrow\infty$\,-\,Hubbard model, one finds $v_{N,c}=2v_c$ 
(see Ref.\ \onlinecite{Schoenhammer05}) so that $v_{J,c}=1/2 v_c\neq v_c$ [see Eq.\ (\ref{defvsc})].

\section{Summary}
\label{summary}

In the present paper we have studied the effect of the different velocities characterizing 
the low-energy physics of a LL on transport through
a one-dimensional, metallic ring of correlated electrons. A magnetic flux
piercing the ring was included and the dependence of the transmission 
probability (linear conductance) on the flux was studied. It was obtained using  
a relation [Eq.~(\ref{transmittance1})] between the one-particle Green function 
of the isolated ring and the transmission probability. 
This becomes exact in the noninteracting limit and serves as an approximation in the 
presence of two-particle interactions. 
Results were averaged over a small energy window around the chemical potential 
just as in Refs. \onlinecite{Jagla} and \onlinecite{Hallberg}. It proved possible to 
improve on an approximation suggested in Ref.\ \onlinecite{Jagla} and to exactly 
calculate the Green functions of the isolated ring for the case of a potential with 
box-like shape in momentum space. Since only the low-energy properties of the 
system are relevant, this restriction to a specially shaped potential should not 
be regarded as severe, in particular because the width of the potential plays 
practically no role for the results.

Characteristic dips appear in the transmission probability as a function of the magnetic flux 
whose number depends on the different velocities in play. According to the ``dominant pole'' approximation 
of Ref.\,\onlinecite{Jagla}, the number of dips for $\phi \in [0,2\pi]$ is given by 
$p \,  v_F/v_s= q \,  v_F/v_c$ (or $p \, v_{J,c}/v_s=q \, v_{J,c}/v_c$ if, as for
systems which are not Galilei invariant, $v_{J,c} \neq v_F$) if 
small odd integers $p$ and $q$ can be found such that $pv_c=qv_s$. For generic repulsive two-particle 
interactions $v_s\le v_c$ and $p$ will be the smaller of the two integers. We have shown that $p$ and $q$ 
must indeed be very small for this prediction for the number of dips to hold. For more generic 
choices of the relevant velocities $v_c$, $v_s$, and $v_{J,c}$, and thus coupling constants,
the number of prominent dips 
in $T(\phi)$ over one interval of periodicity can be estimated to be an integer
close to   $v_{J,c}/v_s$. The wide applicability of this relation was demonstrated.
 
That the charge current velocity $v_{J,c}$ is a relevant parameter determining the flux dependence 
of the linear (charge) conductance through the ring should not come as a surprise. The appearance 
of the velocity of the collective spin excitations $v_s$ can be understood from the structure 
Eqs.\ (\ref{erstegreater}),  (\ref{erstelesser}), (\ref{zweitegreater}), and (\ref{zweitelesser}) 
of the one-particle Green function. Generically the sums appearing in these expressions 
are dominated by the terms in which the integer in front of $v_c$ is zero while the integer 
in front of $v_s$ is greater than zero (but small).  

A simple picture related to the ``dominant pole'' approximation 
suggests that electrons can pass the ring only if spin and charge 
excitations traveling with different velocities (spin-charge separation) can recombine 
at the right contact.  Our results show that this idea of charge and spin ``recombination'' is
of limited usefulness for the understanding of transport through LL rings. 

\section*{Acknowledgments}
We thank K.\ Sch\"onhammer for very valuable discussions and the  
Deutsche Forschungsgemeinschaft (FOR 723) for support.


\begin{thebibliography}{*}
\bibitem{footnote} Throughout this work we assume that $v_c \geq v_s$, which is reasonable 
for repulsive two-particle interactions (see Ref.\ \onlinecite{Schoenhammer05}).
\end{thebibliography}
\end{document}